\documentclass[aps, prd, 11pt, eqsecnum, tightenlines, notitlepage,  nofootinbib, floatfix]{revtex4-1}

\usepackage{feynmp-auto}
\usepackage{amsmath}
\usepackage[utf8x]{inputenc}

\usepackage{bbm}
\usepackage{soul}	
\usepackage{xcolor}

\usepackage{epsfig}
\usepackage{color}
\usepackage{slashed}
\usepackage{comment}
\usepackage{epstopdf}

\usepackage{array}
\usepackage{booktabs}
\usepackage{mathrsfs}
\usepackage[euler]{textgreek}
\usepackage{amsmath}
\usepackage{amsthm}
\usepackage{amsfonts}
\usepackage{amssymb}
\usepackage{graphicx}
\graphicspath{ {Figures/} }
\usepackage[font={small}]{caption}
\usepackage{float}
\usepackage{multirow}
\usepackage[export]{adjustbox}
\usepackage[hang, flushmargin,bottom]{footmisc} 
\usepackage{hyperref}
\definecolor{Color}{rgb}{0.28, 0.24, 0.55}
\hypersetup{
     colorlinks   = true,
     citecolor    =Color,
     linkcolor    =Color,
     urlcolor     =Color
}

\usepackage{placeins}
\usepackage{enumitem}

\usepackage{slashed}

\usepackage[normalem]{ulem}

\DeclareUnicodeCharacter{2212}{-}

\usepackage{threeparttable}
\definecolor{ocre}{rgb}{1,0.6,0.07}
\usepackage{amsfonts}
\usepackage{graphicx}
\usepackage{tcolorbox}
\usepackage{marvosym} 
\usepackage{tikz}
\usepackage{tabulary}
\usepackage{pgfplots}
\usepackage{subfig}
\usepackage{amsmath}
\usepackage{slashed}
\usepackage{mathtools}
\usepackage{nccmath}
\usepackage{verbatim}
\usepackage{color}
\usepackage{amsmath, amsthm, amssymb, amsfonts}
\usepackage{array}
\newcolumntype{P}[1]{>{\centering\arraybackslash}p{#1}}
\usepackage{enumerate}
\usepackage{physics}
\usepackage[font={small, up}]{caption}
\usepackage{tikz}
\usetikzlibrary{"arrows", "automata", "backgrounds", "calendar", "chains", "matrix", "mindmap", "patterns", "petri", "shadows", "shapes.geometric", "shapes.misc", "spy", "trees"}
\graphicspath{{./images/}}
\usepackage{appendix}

\usepackage{tikz}
\usetikzlibrary{arrows,shapes}
\usetikzlibrary{trees}
\usetikzlibrary{matrix,arrows} 				
\usetikzlibrary{positioning}				
\usetikzlibrary{calc,through}				
\usetikzlibrary{decorations.pathreplacing}  
\usepackage{pgffor}							

\usetikzlibrary{decorations.pathmorphing}	
\usetikzlibrary{decorations.markings}
\tikzset{
    vector/.style={decorate, decoration={snake}, draw},
	provector/.style={decorate, decoration={snake,amplitude=2.5pt}, draw},
	antivector/.style={decorate, decoration={snake,amplitude=-2.5pt}, draw},
    fermion/.style={draw=black, postaction={decorate},
        decoration={markings,mark=at position .55 with {\arrow[draw=black]{>}}}},
          fermionD/.style={draw=black, postaction={decorate},
        decoration={markings,mark=at position .75 with {\arrow[draw=black]{>}}}},
    fermionbar/.style={draw=black, postaction={decorate},
        decoration={markings,mark=at position 0.25 with {\arrow[draw=black]{<}}}},
    fermionbar2/.style={draw=black, postaction={decorate},
        decoration={markings,mark=at position 0.25 with {\arrow[draw=black]{>}}}},
    fermionnoarrow/.style={draw=black},
    gluon/.style={decorate, draw=black,
        decoration={coil,amplitude=4pt, segment length=5pt}},
    scalar/.style={dashed,draw=black, postaction={decorate},
        decoration={markings,mark=at position .55 with {\arrow[draw=black]{>}}}},
    scalarbar/.style={dashed,draw=black, postaction={decorate},
        decoration={markings,mark=at position .55 with {\arrow[draw=black]{<}}}},
    scalarnoarrow/.style={dashed,draw=black},
    electron/.style={draw=black, postaction={decorate},
        decoration={markings,mark=at position .55 with {\arrow[draw=black]{>}}}},
	bigvector/.style={decorate, decoration={snake,amplitude=4pt}, draw},
}

\tikzstyle{block} = [draw, rectangle, 
    minimum height=3em, minimum width=6em]

\NewDocumentCommand\semiloop{O{black}mmmO{}O{above}}
{%
\draw[#1] let \p1 = ($(#3)-(#2)$) in (#3) arc (#4:({#4+180}):({0.5*veclen(\x1,\y1)})node[midway, #6] {#5};)
}

\usetikzlibrary{shapes.misc}

\tikzset{cross/.style={cross out, draw=black, minimum size=2*(#1-\pgflinewidth), inner sep=0pt, outer sep=0pt},
cross/.default={1pt}}

\NewDocumentCommand\hello{O{black}mmmO{}O{above}}
{%
\draw[#1] let \p1 = ($(#3)-(#2)$) in (#3) arc (#4:({#4-180}):({0.5*veclen(\x1,\y1)})node[midway, #6] {#5};)
}

\NewDocumentCommand\semiloopI{O{black}mmmO{}O{above}}
{%
\draw[#1] let \p1 = ($(#3)-(#2)$) in (#3) arc (#4:({#4-180}):({0.5*veclen(\x1,\y1)})node[midway, #6] {#5};)
}

\begin{document}

\hspace{5.2in} \mbox{CALT-TH/2021-023}

\title{Scalar Leptoquarks, Baryon Number Violation and Pati-Salam Symmetry}

\author{Clara Murgui and Mark B. Wise}

\affiliation{Walter Burke Institute for Theoretical Physics, California Institute of Technology, Pasadena, CA 91125, USA}

\begin{abstract}
One or more scalar leptoquarks with masses around a few TeV may provide a solution to some of the flavor anomalies that have been observed. We discuss the impact of such new degrees on baryon number violation when the theory is embedded in a Pati-Salam model. The Pati-Salam embedding can suppress renormalizable and dimension-five baryon number violation in some cases. Our work extends the results of Assad, Grinstein, and Fornal who considered the same  issue for vector leptoquarks.

\end{abstract}
\maketitle

\section{Introduction}
The Standard Model (SM) describes a wealth of laboratory data\footnote{There are a few phenomena where that is not true. One must add degrees of freedom (or non-renormalizable operators) to the minimal SM to accommodate neutrino masses (either Majorana or Dirac) and of course there are the well-known cosmological failings.}  within an elegant framework based on the spontaneously broken gauge group ${\rm SU}(3)\times {\rm SU}(2)_L\times {\rm U}(1)_Y$. In the area of flavor physics there are a number of laboratory measurements that point to discrepancies with the SM: muon $g-2$~\cite{Bennett:2006fi,Abi:2021gix}, ${\cal R}_{K^{(*)}}= \text{Br}(B \rightarrow K^{(*)} \mu^+ \mu^-)/ \text{Br}(B \rightarrow K^{(*)} e^+ e^-)$~\cite{Aaij:2014ora,Aaij:2017vbb,Aaij:2019wad,Aaij:2021vac}, and ${\cal R}_{D^{(*)}}= \text{Br}(B \rightarrow D^{(*)} \tau {\bar \nu}_{\tau})/ \text{Br}(B \rightarrow D^{(*)} e (\mu) {\bar \nu}_{e (\mu)})$~\cite{Lees:2012xj,Lees:2013uzd,Aaij:2015yra,Huschle:2015rga,Hirose:2016wfn,Hirose:2017dxl,Aaij:2017uff,Aaij:2017deq,Abdesselam:2019dgh}. Adding leptoquarks (for a review see ref.~\cite{Dorsner:2016wpm}) with masses around a few TeV to the usual SM degrees of freedom has been proposed as an explanation for these discrepancies, see for example~\cite{Gripaios:2009dq,Alonso:2015sja,Calibbi:2015kma,Freytsis:2015qca,Fajfer:2015ycq,Bauer:2015knc,Bhattacharya:2016mcc,Barbieri:2016las,Hiller:2017bzc,Buttazzo:2017ixm,DiLuzio:2017vat,Bordone:2017bld,Becirevic:2018afm,Heeck:2018ntp,Angelescu:2018tyl,Fornal:2018dqn,Aebischer:2019mlg,Popov:2019tyc,Bigaran:2020jil,Cornella:2021sby,FileviezPerez:2021xfw}. Vector and scalar leptoquarks are predicted in Pati-Salam models~\cite{Pati:1974yy}. They are attractive extensions of the SM which have quark-lepton unification based on the ${\rm SU}(4)$ gauge group. 

Treating the SM with minimal particle content as a low-energy effective theory, baryon number violating operators first occur at dimension six and in the Lagrangian these operators are suppressed by a mass scale $\Lambda$ squared~\cite{Weinberg:1979sa,Wilczek:1979hc}. The absence of observed baryon number violation in the laboratory implies that $\Lambda > 10^{15}\text{ GeV}$ (for a review in baryon number violation in various models see ref.~\cite{Nath:2006ut}). Adding TeV mass scalar leptoquarks and diquarks to the SM content can give rise to unacceptably large baryon number violation at the renormalizable and dimension-five levels~\cite{Arnold:2013cva}.
This paper discusses how the pattern of baryon number violation changes if these new scalar degrees of freedom  are embedded in a Pati-Salam model. This issue was considered at the renormalizable level in ref.~\cite{Saad:2017pqj}. Our work is an extension of the work of Assad, Grinstein, and Fornal~\cite{Assad:2017iib}, where baryon number violation mediated by the vector leptoquarks which are associated with the broken generators of the ${\rm SU}(4)$ gauge group was considered.

We consider both the case where the Pati-Salam gauge group is spontaneously broken at a scale much higher than the $\sim 100-1000$ TeV scale and when it is broken at the $\sim 100-1000$ TeV scale.\footnote{It is the limit on the rate for  $K_L \rightarrow \mu^{\pm} e^{\mp}$ that forces the scale to be this high.} In the former case an awkward  tree-level tuning of parameters must be imposed to have one (or more) of the leptoquarks mass much smaller than the symmetry breaking scale. In both the former and latter cases there is the fine-tuning of radiative corrections associated with keeping scalar masses small compared with the Planck scale. However, the same problem exists in the minimal SM for the Higgs boson mass. We will not comment further on these tuning issues in this paper.

In section \ref{sec:LQ} we discuss adding leptoquarks and diquarks to the SM effective theory. We review the implications of this for baryon number violation treating the SM as an effective field theory and using naive dimensional analysis. 

In section \ref{sec:PS} we discuss embedding the SM in Pati-Salam. We discuss the implications that additional light leptoquark(s) have on the expected rates for baryon number violation. Here we treat Pati-Salam as an effective theory with a cutoff that suppresses higher dimension operators invariant under the Pati-Salam gauge group. For Pati-Salam symmetry breaking at a high scale we use the vacuum expectation value (vev) of a field that also gives the right-handed neutrinos a large mass (i.e., type-I seesaw~\cite{Minkowski:1977sc,Yanagida:1979as,GellMann:1980vs,Mohapatra:1979ia}). We find that the scale of Pati-Salam symmetry breaking is constrained, by limits on baryon number violation, to be below $\sim 10^{15}~ {\rm GeV}$. At the same time avoiding strong coupling the scale suppressing higher dimension operators in the Pati-Salam effective theory is constrained to be high, i.e., above $10^{18}~ {\rm GeV}$. In the case of low-scale Pati-Salam breaking we still need the cutoff scale suppressing the higher dimension Pati-Salam invariant operators to be above $\sim 10^{15}$  GeV. We follow ref.~\cite{Perez:2013osa} and generate neutrino masses using the inverse seesaw mechanism~\cite{Mohapatra:1986aw,Mohapatra:1986bd}. In this case we elucidate how the spontaneously broken Pati-Salam symmetry protects the theory from the light scalar leptoquarks giving rise to unacceptably large baryon number violation.

\section{Leptoquarks, Diquarks and Baryon Number Violation}
\label{sec:LQ}
In the SM the renormalizable interactions conserve baryon number (and lepton number) and  the bosons  do not carry any baryon number (or lepton number) charge. Operators that violate baryon number do not occur in the Lagrangian for the SM effective field theory until dimension six. However, this is an accident of the gauge symmetry and the matter content of the SM. Typically, one expects a larger field content when embedding the SM in another theory that is more complete at higher energies (i.e., the UV). For instance, in ${\rm SU}(5)$ grand unified theory a scalar boson that interacts with leptons and quarks at the renormalizable level is predicted together with the SM Higgs boson~\cite{Georgi:1974sy}. 

Leptoquark and diquark scalars can give rise to exotic processes which could be observable in the laboratory. In particular, order ${\cal O}(1-10)$ TeV scalar leptoquarks are popular candidates for explaining the recent anomalies reported in $B$ meson decays that point to violations of lepton flavour universality. Amongst the exotic processes, however, leptoquarks and diquarks can lead to baryon number violation, either at the renormalizable level or by higher dimension operators. Baryon number violation is strongly constrained by laboratory experiments, thereby rendering scalar or vector leptoquarks and diquarks that give rise to renormalizable and dimension-five baryon number violation too heavy to be relevant for the flavor anomalies. 
\begin{table}[h]
    \centering
    \begin{tabular}{|c|c| c | c |}
    \hline
    $({\rm SU}(3), {\rm SU}(2)_L, {\rm U}(1)_Y)_{SM}$ & Renormalizable interactions & Type & B violation \\
    \hline
    \hline
   $(3,3,-1/3)_{SM}$ & $ X_A^\alpha  (Q_L^{\beta} \tau^A Q_L^{\gamma}) \epsilon_{\alpha \beta \gamma} ( \textsc{a})$, $X_A^\alpha ((Q_{L}^\dagger)_\alpha  \tau^A L_L^\dagger) $  & M & \text{dim 4} \\
     $ (3,1,-4/3)_{SM}$  & $X^\alpha ((d_{R}^\dagger)_\alpha e_R^\dagger)$, $X^\alpha (u_R^\beta u_R^\gamma)  \epsilon_{\alpha \beta \gamma} \ (\textsc{a})$ & M & dim 4 \\
  $(3,1,-1/3)_{SM}$ & $X^\alpha (Q_L^\beta  Q_L^\gamma)  \epsilon_{\alpha \beta \gamma} (\textsc{s})$, $X^\alpha ((Q_{L}^\dagger)_\alpha L_L^\dagger)$, $X^\alpha (u_R^\beta d_R^\gamma)  \epsilon_{\alpha \beta \gamma}$, $X^\alpha ((u_{R}^\dagger)_\alpha e_R^\dagger) $ & M & dim 4 \\
   $(3,2,7/6)_{SM}$ & $ X^\alpha ((Q_{L}^\dagger)_\alpha  e_R ) $, $X^\alpha  (L_L  (u^\dagger_{R})_\alpha) $ & LQ & dim 5 \\
  $(3,2,1/6)_{SM}$ &  $X^\alpha (L_L (d^\dagger_{R})_\alpha) $ & LQ & dim 5 \\
  $(3,1,2/3)_{SM} $ & $X^\alpha (d_R^\beta d_R^\gamma)  \epsilon_{\alpha \beta \gamma} (\textsc{a})$ & DQ & dim 5 \\
  $(\bar 6,3,-1/3)_{SM}$ & $ X^A_{\alpha \beta} (Q_L^\alpha  \tau_A Q_L^\beta) \ (\textsc{s})$ & DQ & dim 5\\ 
  $(\bar 6,1,-1/3)_{SM}$ & $X_{\alpha \beta} (u_R^\alpha d_R^\beta)$, $ X_{\alpha \beta} (Q_L^\alpha  Q_L^\beta) \ (\textsc{a})$ & DQ & dim 5 \\
  $(\bar 6,1,2/3)_{SM}$ & $X_{\alpha \beta} (d_R^\alpha d_R^\beta) \ (\textsc{s})$ & DQ & dim 5\\
$(\bar 6,1,-4/3)_{SM}$ & $ X_{\alpha \beta} (u_R^\alpha  u_R^\beta) \ (\textsc{s})$ & DQ & $>$ dim 6 \\
    \hline
\end{tabular}
    \caption{Scalar leptoquarks and diquarks. In the first column, their quantum numbers under the SM gauge group are listed. In the second column, we list their renormalizable interactions with matter. Between parenthesis right after the Yukawa interaction we indicate, for those which enjoy some flavor symmetry, whether the flavor indices are symmetric ($\textsc{s}$) or antisymmetric ($\textsc{a}$). In the third column we classify them as pure diquark (DQ), pure leptoquark (LQ) or mixed (M) scalars, as described in the text. In the last column we specify at which dimension in the SM effective field theory baryon number violation occurs.}
\label{tab:SLQ}
\end{table}

In this paper we use both left and right-handed two-component fields and spinors denoted by subscripts L and R, respectively. We do not explicitly display Lorentz and ${\rm SU}(2)_L$ gauge indices. We also do not explicitly display the two-index antisymmetric tensors used to contract those indices. In addition flavor indices are often not displayed explicitly.

In table~\ref{tab:SLQ} we list all possible leptoquark and diquark scalars able to interact at the renormalizable level with matter,\footnote{We are assuming one new boson at a time. If two or more are present, new interactions could arise through the mixing terms in the scalar potential.} according to Lorentz invariance and the SM gauge symmetry, that carry a non-zero baryon number charge for their individual Yukawa interactions to conserve baryon number. In the second column we show the explicit interactions with matter at the renormalizable level. In the third column we classify the type of the scalar:
\begin{itemize}
    \item Pure diquark (DQ): The scalar $X$ has only renormalizable interactions with a pair of quarks, i.e. $\!X$ has a non-zero baryon number charge.
    \item Pure leptoquark (LQ): The scalar $X$ only has interactions where both quarks and leptons are involved at the renormalizable level, i.e. $\!X$ has non-zero baryon and lepton number charges.
    \item Mixed (M): The scalar $X$ has interactions with a pair of quarks as well as with both leptons and quarks. No baryon or lepton number charge can be assigned to the scalar in this case.
\end{itemize}

\begin{figure}[t]
\begin{equation*}
\begin{gathered}
\subfloat{
\resizebox{.5\textwidth}{!}{%
\begin{tikzpicture}[line width=1.5 pt,node distance=1 cm and 1 cm]
\coordinate[](bola);
\coordinate[left= 1.8 cm of bola, label=left:$p$](aux);
\coordinate[above left = of bola, label=left:$u_R$] (uR);
\coordinate[left = of bola, label=left:$d_L$](dL);
\coordinate[below left = of bola, label=left:$u_L$] (uL);
\coordinate[above right = of bola](vX);
\coordinate[below = 0.35cm of vX, label=below:$\,\,\, c_R\text{, }t_R$](aux1);
\coordinate[right = of vX](vX2);
\coordinate[right = of vX2,label=right:$e_R$](eR);
\coordinate[left= 1.5cm of eR,label=above:$X$](X);
\coordinate[below right = of vX2, label=right:$d_R$](dR);
\coordinate[below right = of bola](pW);
\coordinate[below = of dR,label=right:$ d_R$](dLc);
\coordinate[above right = 1.2 cm of dLc,label=below:$ \quad \quad \quad \ \pi^0$](aux3);
\coordinate[above=0.2cm of pW,label=$W$](aux4);
\draw[fermion] (uR) -- (vX);
\draw[scalar] (vX2)--(vX);
\draw[fermion] (uL) -- (pW);
\draw[fermion](dL)--(bola);
\draw[fermion] (bola) -- (vX);
\draw[vector] (pW) -- (bola);
\draw[fermion] (eR)--(vX2);
\draw[fermion] (dR)--(vX2);
\draw[fermion](pW)--(dLc);
\draw[fill=black](vX) circle (0.05cm);
\draw[fill=black](vX2) circle (0.05cm);
\draw[fill=black](pW) circle (0.05cm);
\draw[fill=black] (bola) circle (0.05cm);
   \draw[fill=gray!40!white,draw=black] (-1.75,0) ellipse (0.1cm and 1cm);
      \draw[fill=gray!40!white,draw=black] (4,-0.5) ellipse (0.1cm and 0.6cm);
\end{tikzpicture}}\label{fig:panelA}}
\quad \, \,
\resizebox{.38\textwidth}{!}{
\begin{tikzpicture}[line width=1.5 pt,node distance=1 cm and 1.5 cm]
\coordinate[label=left:$d_R$](d1);
\coordinate[below right = 1.5 cm of d1](v1a);
\coordinate[below left =1.5 cm  of v1a,label=left:$u_R$](l);
\coordinate[right =  of v1a](v2a);
\coordinate[right = 0.7 cm of v1a, label=below:$\Phi_3$](vaux);
\coordinate[above=0.7cm of v1a](vaux2);
\coordinate[right=0.3cm of vaux2,label=above:$\langle H \rangle$](vHiggs);
\coordinate[above right = 1.5 cm of v2a, label=right:$d_R$](d2);
\coordinate[below right = 1.5 cm of v2a,label=right:$L_L$](d3);
\draw[fermion](d1)--(v1a);
\draw[fermion] (l)--(v1a);
\draw[scalar](v1a)--(v2a);
\draw[fermion] (d2)--(v2a);
\draw[fermion] (v2a)--(d3);
\draw[scalarnoarrow](vHiggs)--(v1a);
\draw[fill=black] (v2a) circle (.05cm);
\filldraw[fill=gray!40!white, draw=black] (v1a) circle (.2cm);
\end{tikzpicture}}
\end{gathered}
\end{equation*}
\begin{equation*}
\quad \quad \ \ 
\begin{gathered}
\resizebox{.42\textwidth}{!}{
\begin{tikzpicture}[line width=1.5 pt,node distance=1 cm and 1.5 cm]
\coordinate(vcentral);
\coordinate[above left =  1.5cm of vcentral](v1);
\coordinate[above right =  1.5cm of vcentral](v2);
\coordinate[below right = 1.5cm of vcentral,label=right:$\, e_R$](v3);
\coordinate[below left = 1.5cm  of vcentral,label=left:$d_R \  $](v4);
\coordinate[below =  0.9 cm of v1,label=$\ \Phi_3$](v1aux);
\coordinate[below =  0.9 cm of v2,label=$\! \Phi_4$](v2aux);
\coordinate[left= 0.75 cm of v1,label=left:$d_R$](f1a);
\coordinate[above = 0.75cm of v1,label=above:$L_L$](f1b);
\coordinate[above = 0.75cm of v2,label=above:$Q_L\text{, }u_R$](f2a);
\coordinate[right = 0.75 cm of v2,label=right: $e_R\text{, }L_L$](f2b);
\draw[fermion](f1a)--(v1);
\draw[fermion](v1)--(f1b);
\draw[fermion](f2a)--(v2);
\draw[fermion](v2)--(f2b);
\draw[scalar] (vcentral) -- (v1);
\draw[scalar] (v2) -- (vcentral);
\draw[fermion] (v3) -- (vcentral);
\draw[fermion] (v4) -- (vcentral);
\filldraw[fill=gray!40!white, draw=black] (vcentral) circle (.2cm);
\draw[fill=black] (v1) circle (.05cm);
\draw[fill=black] (v2) circle (.05cm);
\end{tikzpicture}}
\end{gathered} 
\quad \quad \quad  
\begin{gathered}
\resizebox{.38\textwidth}{!}{
\begin{tikzpicture}[line width=1.5 pt,node distance=1 cm and 1.5 cm]
\coordinate(vcentral);
\coordinate[above left =  1.5cm of vcentral](v1);
\coordinate[above right =  1.5cm of vcentral](v2);
\coordinate[below right = 1.5cm of vcentral,label=right:$\, d_R$](v3);
\coordinate[below left = 1.5cm  of vcentral,label=left:$d_R \  $](v4);
\coordinate[below =  0.9 cm of v1,label=$\ X$](v1aux);
\coordinate[below =  0.9 cm of v2,label=$\! X$](v2aux);
\coordinate[left= 0.75 cm of v1,label=left:$d_R$](f1a);
\coordinate[above = 0.75cm of v1,label=above:$u_R$](f1b);
\coordinate[above = 0.75cm of v2,label=above:$d_R$](f2a);
\coordinate[right = 0.75 cm of v2,label=right: $u_R$](f2b);
\draw[fermion](f1a)--(v1);
\draw[fermion](f1b)--(v1);
\draw[fermion](f2a)--(v2);
\draw[fermion](f2b)--(v2);
\draw[scalar] (vcentral) -- (v1);
\draw[scalar] (vcentral) -- (v2);
\draw[fermion] (v3) -- (vcentral);
\draw[fermion] (v4) -- (vcentral);
\filldraw[fill=gray!40!white, draw=black] (vcentral) circle (.2cm);
\draw[fill=black] (v1) circle (.05cm);
\draw[fill=black] (v2) circle (.05cm);
\end{tikzpicture}}
\end{gathered} 
\end{equation*}
\caption{On the top panels, diagrams for $\Delta B = -\Delta L = -1$ processes. On the bottom panels, diagram for $\Delta B = - \Delta L = -1$ (bottom-left panel) and diagram for $\Delta B = -2, \Delta L = 0$ (bottom-right panel) processes.}
\label{fig:FG}
\end{figure}
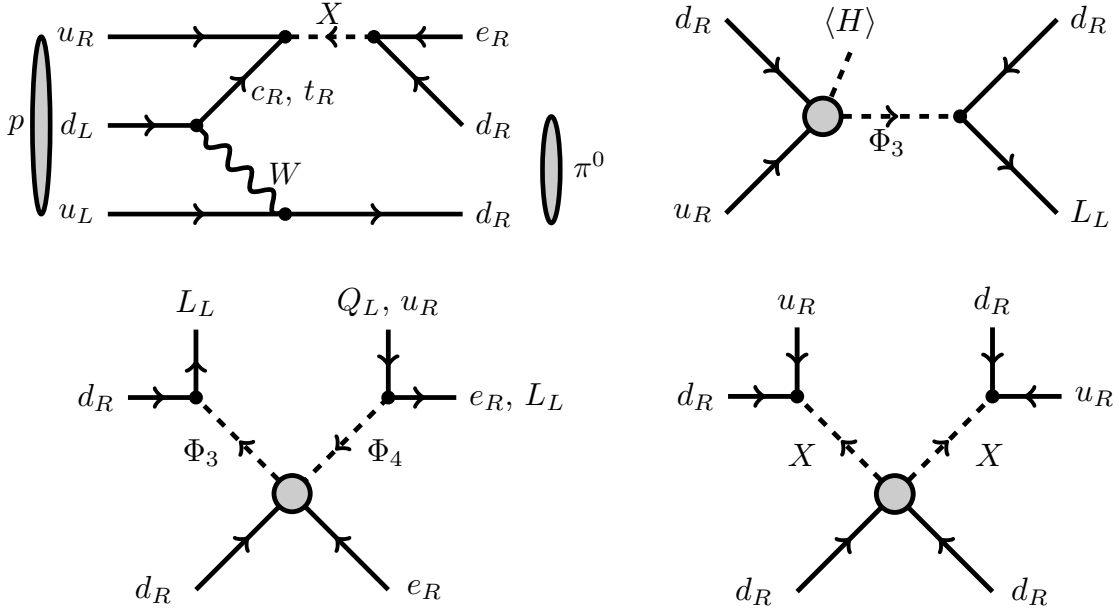

The mixed scalars violate baryon number at the renormalizable level due to the simultaneous presence of diquark and leptoquark couplings. In general, they contribute to $|\Delta B|=1$ processes through tree-level $X$ exchange. Consistency with laboratory experiments requires (assuming order one Yukawa couplings) that $M_X \gtrsim 10^{15}$ GeV. Exceptions of this case are the mixed scalars whose diquark coupling with up quarks is antisymmetric in the flavor indices. For example, the scalar $X \sim (3,1,-4/3)_{SM}$ whose renormalizable interactions with matter are
\begin{equation}
-{\cal L}  \supset Y_\text{DQ} (u_R^\alpha u_R^\beta)  X^\gamma \epsilon_{\alpha \beta \gamma} + Y_\text{LQ} (d_R^\alpha e_R) X^\dagger_\alpha + \text{h.c.}~.
\end{equation} 
Here and later in this section the Yukawa matrices are in the mass eigenstate basis. 
In this case the proton can only decay by an additional exchange of a $W$ gauge boson~\cite{Dong:2011rh,Arnold:2012sd,Dorsner:2012nq} as the diagram in the top-left panel from fig.~\ref{fig:FG} shows. Following the calculation in ref.~\cite{Dorsner:2012nq} for the decay rate of the proton via the most experimentally constrained mode $p \to \pi^0 e^+$\footnote{In eq.~\eqref{eq:Wexchange} we have updated the proton lifetime to the recent limit $\tau(p \to \pi^0 e^+) > 2.4 \times 10^{34}$ years~\cite{Takenaka:2020vqy} by the Super-Kamiokande collaboration.} implies
\begin{equation}
 M_X \gtrsim 3 \times 10^{12} \text{ GeV }\left|\sum_{Q=2,3}V_\text{CKM}^{Q1}Y_\text{DQ}^{1Q} Y_\text{LQ}^{11}\left(\frac{\text{GeV}}{m_Q}\right)\right |^{1/2}.
 \label{eq:Wexchange}
\end{equation}
 In the above expression the superscripts correspond to the entries of the matrix in the flavor space. We will adopt this convention in the equations where the flavor of fermion fields is explicitly displayed. Assuming order one Yukawa couplings, from the above equation one obtains that $M_X \gtrsim 10^{12}$ GeV, ruling out TeV scale mixed scalars.

Pure leptoquark scalars can be given a definite charge so that baryon number is simultaneously conserved in all the allowed renormalizable interactions.\footnote{Naively one might have thought that the $(3,2,1/6)_{SM}$ leptoquark gives rise to renormalizable baryon number violation through the operator $\epsilon_{\alpha \beta \gamma}\epsilon_{ij}X^{\alpha i}X^{\beta j}X^{\gamma k}H^{\dagger}_k$. Here we have explicitly displayed the ${\rm SU}(2)_L$ indices $i,j,k$ which take on values 1 and 2. Expanding the ${\rm SU}(2)_L$ contractions the operator contains at least two $X$'s with the same ${\rm SU}(2)_L$ index value and  so it vanishes because of the antisymmetry of the $\epsilon_{\alpha \beta \gamma}$ used to contract the $X$ color indices.} There are only two scalars in table~\ref{tab:SLQ} satisfying this property, $\Phi_3 \sim (\bar 3,2,-1/6)_{SM}$ and $\Phi_4 \sim (3,2,7/6)_{SM}$.\footnote{The nomenclature $\Phi_3$ and $\Phi_4$ used for these scalar leptoquarks will be motivated in the following section.} However, as the authors in ref.~\cite{Arnold:2013cva} pointed out, baryon number can still be violated by operators of dimension five, which generate a diquark coupling as listed below:
\begin{eqnarray}
 \frac{\epsilon_{\alpha \beta \gamma}}{\Lambda} (d_R^\alpha u_R^\beta) (\Phi_3^\dagger)^\gamma H^\dagger  , \quad
\frac{\epsilon_{\alpha \beta \gamma}}{\Lambda} (Q_L^\alpha Q_L^\beta) (\Phi_3^\dagger)^\gamma H^\dagger ,\quad
 \frac{\epsilon_{\alpha \beta \gamma}}{\Lambda}  (d_R^\alpha d_R^\beta) (\Phi_3^\dagger)^\gamma H  , \quad \frac{\epsilon_{\alpha \beta \gamma}}{\Lambda} (d_R^\alpha d_R^\beta) \Phi_4^\gamma H^\dagger.
 \label{eq:dim5SLQ}
\end{eqnarray}
The top-right panel of fig.~\ref{fig:FG} shows the diagram contributing to $\Delta B = - \Delta L = -1$ processes corresponding to the first operator above. In eq.~\eqref{eq:dim5SLQ} and elsewhere the hermitian conjugate partners are understood to also be present. In ref.~\cite{Arnold:2013cva} it is explicitly shown that experimental bounds on nucleon decays rule out the two pure leptoquarks with TeV scale masses even when the cutoff $\Lambda$ is equal to the Planck scale (note that any coupling constants are absorbed into $\Lambda$ in eq.~\eqref{eq:dim5SLQ}).

The following dimension-five operators also contribute to nucleon decay,
\begin{equation}
    \frac{\epsilon_{\alpha \beta \gamma}}{\Lambda} (d_R^\alpha e_R) (\Phi_3^\dagger)^\beta \Phi_4^\gamma , \quad \frac{\epsilon_{\alpha \beta \gamma}}{\Lambda} (Q_L^\alpha L_L) (\Phi_3^\dagger)^\beta (\Phi_3^\dagger)^\gamma , \quad \frac{\epsilon_{\alpha \beta \gamma}}{\Lambda} (u_R^\alpha e_R) (\Phi_3^\dagger)^\beta (\Phi_3^\dagger)^\gamma.
    \label{eq:dim5SLQextra}
\end{equation}
as displayed in the bottom-left panel of fig.~\ref{fig:FG} for the first of the above operators.
However they give rise to rates that are suppressed compared to those that follow from the operators in eq.~\eqref{eq:dim5SLQ}. 

The same argument applies to the pure diquarks. Although they seem harmless at the renormalizable level, leptoquark couplings could arise in dimension-five operators leading to baryon number violating processes. For instance, the pure diquark $X \sim (3,1,2/3)_{SM}$ has the following leptoquark couplings at dimension five, 
\begin{equation}
  \frac{1}{\Lambda}  H^\dagger (L_L (d_{R}^\dagger)_\alpha) X^\alpha,~~\frac{1}{\Lambda}H((u_R^\dagger)_\alpha L_L)X^\alpha,~~{\rm and}~~ \frac{1}{\Lambda}  ((Q_{L}^\dagger)_\alpha e_R) H X^\alpha ,
\end{equation}
which lead to $\Delta B = -\Delta L = -1$ processes similar to the top-right panel of fig.~\ref{fig:FG}. Analogously to the pure leptoquark cases, these processes rule out a TeV scale $X$ mass for Yukawa couplings $Y_\text{DQ} \sim {\cal O}(1)$ even when the cutoff is equal to the Planck mass~\cite{Arnold:2013cva}. 
 
 The scalar diquarks that are in the symmetric $6$ representation of ${\rm SU}(3)$ do not give rise to baryon number violating processes at the renormalizable level. At dimension five $|\Delta B| = 1$ processes do not occur, however $|\Delta B| =2$ processes can be present. To illustrate this consider the pure diquark $X \sim (\bar 6, 1, -1/3)_{SM}$. Through the dimension-five operator
\begin{equation}
  \frac{1}{\Lambda} (X^\dagger)^{\alpha \alpha'} (X^\dagger)^{\beta \beta'} (d_R^\gamma d_R^{\gamma'}) \epsilon_{\alpha \beta \gamma} \epsilon_{\alpha' \beta' \gamma'},
\end{equation}
and the renormalizable diquark interactions listed in table~\ref{tab:SLQ}, the following term in the effective Hamiltonian (amongst others\footnote{Other operators with $\Delta B = -2$ are also generated due to the two possible diquark couplings of $X\sim (\bar 6,1,-1/3)_{SM}$. In the above example we have not considered the interaction $(Q_L^\alpha Q_L^\beta) X_{\alpha \beta} $ because it is less constrained experimentally due to the antisymmetry in the flavor indices.}) is generated once the scalar $X$ is integrated out,
\begin{equation}
  {\cal H}_\text{eff} \supset  \frac{1}{\Lambda} \frac{1}{M_X^4} (d_R^\alpha u_R^{\alpha'}) (d_R^\beta u_R^{\beta'}) (d_R^\gamma d_R^{\gamma'}) \epsilon_{\alpha \beta \gamma} \epsilon_{\alpha' \beta' \gamma'}+ \text{h.c.}~.
\end{equation}
This is illustrated in the bottom-right panel of fig.~\ref{fig:FG}. Such interaction leads to $|\Delta B| = 2$ transitions, for example $n{\rm -}\bar n$ oscillations. Following ref.~\cite{Arnold:2012sd} the bound on the mass of $X$ coming from the experimental limit on $n{\rm -}\bar n$ oscillations in this case is
\begin{equation}
    M_X \gtrsim 300 \text{ GeV}  \left(\frac{M_{PL}}{\Lambda}\right)^{1/4}.
\end{equation}
Therefore, these scalar diquarks in the symmetric representation of ${\rm SU}(3)$ can be at the TeV scale and be consistent with laboratory constraints on baryon number violation for a cutoff in the allowed range $10^{15} \text{ GeV} < \Lambda < M_{PL} = 10^{19} \text{ GeV}$.

Here we have assumed only one light scalar. Other baryon number violating processes are possible if several are light.

\FloatBarrier

\section{Embedding in Pati-Salam}
\label{sec:PS}

The take-home message from the previous section is that TeV scale (color triplet) leptoquarks are generically ruled out by laboratory constraints on baryon number violating processes. In this section, we point out that this need not be the case if the SM is embedded in a larger symmetry group. We will show that  embedding  the SM in the minimal theory for quark-lepton unification {\it à la} Pati-Salam~\cite{Pati:1974yy}, based on the gauge symmetry group~\cite{Smirnov:1995jq,Perez:2013osa}, 
$$ {\rm SU}(4) \otimes {\rm SU}(2)_L \otimes {\rm U}(1)_R$$
protects the Pati-Salam effective theory from unacceptably large baryon number violation at the renormalizable level and by higher dimension operators. 

The matter content we use is as follows. Each family of the SM fermions and a right-handed neutrino are collected in three representations,\footnote{We have suppressed the lepton field in the subscript used for the right-handed Pati-Salam representations to avoid confusion with the field-stength tensor $F_{\mu \nu}$ in the $F_u$ case.}
\begin{equation}
\begin{split}
&F_{d} = (d_R^r,d_R^b,d_R^g, e_R) \sim (4,1,-1/2)_{PS}, \quad F_{u}  = (u_R^r, u_R^b, u_R^g, \nu_R) \sim (4,1,1/2)_{PS},\\
& \qquad \qquad \qquad \quad \text{ and } \quad F_{QL} = (Q_L^r, Q_L^b, Q_L^g, L_L) \sim (4,2,0)_{PS},
\end{split}
\end{equation}
where the three colors of QCD (red ($r$), blue ($b$) and green ($g$)) are explicitly displayed to illustrate that leptons are treated as the fourth color in the matter representations. Only the representations containing right-handed fields are charged under ${\rm U}(1)_R$. As before, color indices which take values $r,b,g$ (or equivalently $1,2,3$) will be represented by indices $\alpha, \beta, \gamma , \ldots$~. ${\rm SU}(4)$ indices that go over $1,2,3,4$ will be denoted by $A,B,C, \ldots$~. In this section we will denote quantum numbers under the Pati-Salam gauge group, $({\rm SU}(4),{\rm SU}(2)_L,{\rm U}(1)_R)$, with the subscript $PS$, and quantum numbers under the SM gauge group, (${\rm SU}(3),{\rm SU}(2)_L,{\rm U}(1)_Y$), with the subscript $SM$.

Apart from the matter content, the theory requires scalars in order to reproduce the experimental observations. Firstly, a scalar non-singlet ${\rm SU}(4)$ representation with a neutral component under the SM gauge charges is needed to break ${\rm SU}(4)$ down to ${\rm SU}(3)$. For instance, a scalar representation with quantum numbers $\chi \sim (4,1,1/2)_{PS}$ or $\Delta \sim (\overline{ 10}, 1, -1)_{PS}$ can do the job with $\langle \chi^A \rangle=\delta
^{A4}v_\chi/\sqrt{2}$ or $\langle \Delta_{AB} \rangle = \delta_{A4}\delta_{B4} v_\Delta / \sqrt{2}$, respectively.
Each of these vevs breaks ${\rm SU}(4)\otimes {\rm U}(1)_R$ to ${\rm SU}(3) \otimes {\rm U}(1)_Y$ and defines the hypercharge of the SM fields. The hypercharge is given by a combination of the charge $R$ and the non-singular diagonal generator of ${\rm SU}(4)$, i.e.
\begin{equation}
  Y = R +\frac{\sqrt{6}}{3}T_4 , \quad \text{ where }   \quad T_4 = \frac{1}{2\sqrt{6}} \text{diag}(1,1,1,-3).
\end{equation}

Secondly, the SM Higgs field $H\sim (1,2,1/2)_{PS}$ is added to spontaneously break the electroweak symmetry.
At this stage the theory predicts equal masses for down quarks and charged leptons, as well as up quarks and neutrinos. A third scalar field in the adjoint representation of ${\rm SU}(4)$ is added to achieve realistic fermion masses,
\begin{equation}
 \Phi_{15} = \begin{pmatrix} \Phi_8 & \Phi_3 \\ \Phi_4 & 0 \end{pmatrix} + H_2 T_4\sim (15,2,1/2)_{PS}.
\end{equation}
It contains a second Higgs doublet, $H_2 \sim (1,2,1/2)_{SM}$, whose coupling to matter distinguishes leptons from quarks. The Yukawa couplings in the Lagrangian, 
\begin{equation}
   - {\cal L} = Y_1  H (F_{QL}^A(F_{u}^\dagger)_A) + Y_2  (\Phi_{15})_A^B (F_{QL}^A(F_{u}^\dagger)_B)  + Y_3  H^\dagger (F_{QL}^A  (F_{d}^\dagger)_A) + Y_4  (\Phi_{15}^\dagger)^B_A (F_{QL}^A (F_{d}^\dagger)_B),
\end{equation}
generate the following mass matrices for the SM fermions after the spontaneous symmetry breaking of the electroweak symmetry,
\begin{equation}
\begin{split}
    M_u & = Y_1 \frac{v_H}{\sqrt{2}} + \frac{1}{2\sqrt{6}}Y_2 \frac{v_\Phi}{\sqrt{2}}, \qquad \qquad M_d = Y_3 \frac{v_H}{\sqrt{2}} + \frac{1}{2\sqrt{6}}Y_4 \frac{v_\Phi}{\sqrt{2}},\\
    M_\nu^{\rm Dirac} & = Y_1 \frac{v_H}{\sqrt{2}} - \frac{3}{2\sqrt{6}} Y_2 \frac{v_\Phi}{\sqrt{2}}, \qquad \qquad M_e = Y_3 \frac{v_H}{\sqrt{2}} - \frac{3}{2\sqrt{6}} Y_4 \frac{v_\Phi}{\sqrt{2}},
 \end{split}
 \label{eq:masses}
\end{equation}
where $\langle H^j \rangle = \delta^{j2} v_H / \sqrt{2} \, $ and $ \langle H_2^j \rangle = \delta^{j2}  \, v_\Phi / \sqrt{2} $. In addition to the second Higgs doublet, a color octet Higgs field $\Phi_8 \sim (8,2,1/2)_{SM}$ and the scalar leptoquarks $\Phi_3 \sim (\bar 3, 2, -1/6)_{SM}$ and $\Phi_4 \sim (3,2,7/6)_{SM}$ are also contained in the $\Phi_{15}$ representation. The interaction terms of the leptoquarks with matter are,
\begin{equation}
-{\cal L} \supset Y_2 (Q_L^\alpha   \nu_R^\dagger)\Phi_{3\alpha} + Y_2 (L_L   (u_R^\dagger)_\alpha)\Phi_4^\alpha + Y_4 (Q_L^\alpha  e_R^\dagger)(\Phi_4^\dagger)_\alpha + Y_4 (L_L  (d_R^\dagger)_\alpha)(\Phi_3^\dagger)^\alpha + \text{h.c.}~,
\label{eq:yukawaPhi3and4}
\end{equation}
which do not violate baryon (or lepton) number if baryon (lepton) number charges $Q_B (\Phi_3) = -1/3$ and $Q_B(\Phi_4) = 1/3$ ($Q_L(\Phi_3)=1$ and $Q_L(\Phi_4)=-1$) are assigned to them. Under these charge assignments, baryon and lepton number are also conserved by the renormalizable interactions in the scalar potential. Following the classification introduced in section~\ref{sec:LQ}, $\Phi_3$ and $\Phi_4$ are the pure leptoquarks of table~\ref{tab:SLQ}.

The gauge sector of this theory also does not violate baryon number at the renormalizable level. The massive gauge fields associated with the broken Pati-Salam generators transform under the SM gauge group as $V\sim (3,1,2/3)_{SM}$. As Assad, Fornal and Grinstein pointed out in ref.~\cite{Assad:2017iib}, adding this vector leptoquark to the SM effective field theory gives rise to baryon number violation at the dimension-five level and if its embedded in Pati-Salam this is absent. We shall see that in the Pati-Salam effective theory the gauge bosons associated with the broken generators first give rise to baryon number violation at dimension seven. Broken gauge generators however do give rise to meson leptonic decays such as $K_L \to \mu^\mp e^\pm$ which set a bound on their mass of $M_V \gtrsim 10^6$ GeV~\cite{Valencia:1994cj,Smirnov:2007hv}, up to possible suppressions coming from the unitary matrices introduced to diagonalize the fermion mass matrices.

This is the field content composing of the minimal theory based on Pati-Salam symmetry. However this theory would require a very awkward tree-level tuning of parameters to get small neutrino masses. In this paper we will add fields that give neutrino masses either by the type-I or inverse seesaw mechanisms. Note from eq.~\eqref{eq:masses} that the Yukawa couplings $Y_2$ and $Y_4$ are the ones that determine the coupling of the leptoquarks to the fermions. Either the Yukawa coupling $Y_1$ or $Y_2$ (but not both) could be very small and still reproduce the correct fermion masses. The Yukawa couplings $Y_3$ and $Y_4$ are constrained by the masses of the down quarks and charged leptons. For simplicity in this paper we assume that the matrices that diagonalize the charged fermion mass matrices are close to the identity. For example, 
\begin{equation}
    Y_4 = \frac{\sqrt{3}}{v} (M_d-M_e) \sim \frac{\sqrt{3}}{v}M_d^\text{diag}.
    \label{eq:texture}
\end{equation}
In the above expression and later in this section we assume that $v = v_H = v_\Phi = 246/\sqrt{2} \text{ GeV}$.

\begin{figure}[t]
\begin{equation*}
\hspace{-0.5cm}
\begin{gathered}
 \resizebox{.45\textwidth}{!}{%
\begin{tikzpicture}[line width=1.5 pt,node distance=1 cm and 1.5 cm]
\filldraw[fill=gray!20!white, draw=none] (0,4) rectangle (4,4.5);
\node (A) at (0, 0) {};
\node (B) at (0, 5) {\scriptsize ${\rm Energy}$};
\node (C) at (2, 0.2) {\scriptsize Leptoquark(s)};
\node (D) at (-0.75, 0.5) {\scriptsize few TeV};
\node (E) at (2, 2.77) {\tiny $\gtrsim 10^{13} \text{ GeV}$ (type-I seesaw)};
\node (F) at (2, 3.5) {\scriptsize ${\rm SU}(4) \otimes {\rm SU}(2)_L \otimes {\rm U}(1)_R$};
\node (G) at (2, 1.75) {\scriptsize ${\rm SU}(3) \otimes {\rm SU}(2)_L \otimes {\rm U}(1)_Y$};
\node (H) at (2, 0.7) {\tiny $\lesssim 10 \text{ TeV}$ (flavor anomalies)};
\node (I) at (-0.5, 3) {\scriptsize $ v_\Delta $};
\node (J) at (-0.5, 4) {\scriptsize $\Lambda_{PS}$};
\node (I) at (2, 4.28) {\scriptsize $\lesssim M_{PL}$};
\node (K) at (1.9, 5.75) {\scriptsize {\bf \text{\underline{High-scale Pati-Salam breaking ($\Delta$)}}}};
\draw[cyan] (0,0.5) -- (4,0.5);
\draw[ocre] (0,3) -- (4,3);
\draw[violet] (0,4) -- (4,4);
\draw[->, to path={-| (\tikztotarget)}]
  (A) edge (B);
\end{tikzpicture}}
\end{gathered}
\quad \quad \quad 
\begin{gathered}
 \resizebox{.45\textwidth}{!}{%
\begin{tikzpicture}[line width=1.5 pt,node distance=1 cm and 1.5 cm]
\filldraw[fill=gray!20!white, draw=none] (0,4) rectangle (4,4.5);
\node (A) at (0, 0) {};
\node (B) at (0, 5) {\scriptsize ${\rm Energy}$};
\node (C) at (2, 0.2) {\scriptsize Leptoquark(s)};
\node (D) at (-0.75, 0.5) {\scriptsize few TeV};
\node (E) at (2.1, 3) {\tiny $\gtrsim 10^{15} \text{ GeV}$ (dim 6 $\Delta B\!  = \! -1$)};
\node (F) at (2, 2.2) {\scriptsize ${\rm SU}(4) \otimes {\rm SU}(2)_L \otimes {\rm U}(1)_R$};
\node (G) at (2, 0.7) {\scriptsize ${\rm SU}(3) \otimes {\rm SU}(2)_L \otimes {\rm U}(1)_Y$};
\node (I) at (-0.5, 1.3) {\scriptsize $ v_\chi $};
\node (J) at (-0.5, 3.25) {\scriptsize $\Lambda_{PS}$};
\node (I) at (2, 4.28) {\scriptsize $\lesssim M_{PL}$};
\node (K) at (1.9, 5.75) {\scriptsize {\bf \text{\underline{Low-scale Pati-Salam breaking ($\chi$)}}}};
\node (L) at (2, 1.1) {\tiny $\gtrsim 10^3 \text{ TeV}$ ($K_L \to \mu^\pm e^\mp$)};
\draw[cyan] (0,0.5) -- (4,0.5);
\draw[ocre] (0,1.3) -- (4,1.3);
\draw[violet] (0,3.25) -- (4,3.25);
\draw[->, to path={-| (\tikztotarget)}]
  (A) edge (B);
\end{tikzpicture}}
\end{gathered}
\end{equation*}
\caption{Energy scales for the two breaking scenarios of the Pati-Salam gauge group. On the left panel, the high-scale Pati-Salam breaking induced by the vev  of $\Delta \sim (\overline{10},1,-1)_{PS}$ described in section~\ref{subsec:PSHigh}. Neutrinos acquire mass through the type-I seesaw mechanism in this scenario. On the right panel, the low-scale Pati-Salam breaking induced by the vev of $\chi \sim (4,1,1/2)_{PS}$ described in section~\ref{subsec:PSLow}. Neutrinos acquire mass through the inverse seesaw mechanism in this scenario.} 
\label{fig:diagramscales}
\end{figure}
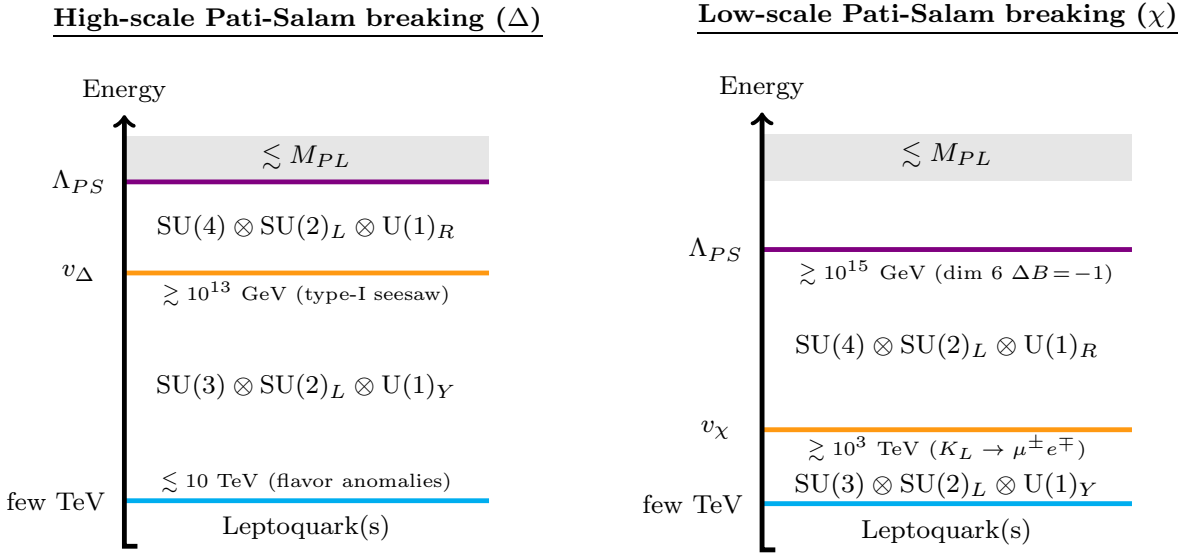

Hereafter we distinguish two possible scenarios for the breaking of ${\rm SU}(4)$: the case where $v_\chi \sim  100-1000\text{ TeV}$, and the case where the ${\rm SU}(4)$ breaking occurs well above $1000$ TeV. We call these respectively low-scale Pati-Salam breaking and high-scale Pati-Salam breaking. See fig.~\ref{fig:diagramscales} for a summary of the different energy scales involved in these scenarios. 

In either the high or low-scale Pati-Salam symmetry breaking scenarios baryon number violation occurs at the dimension-six level through the following terms in the Pati-Salam effective field theory,
\begin{equation}
    {1 \over \Lambda_{PS}^2}\epsilon_{ABCD}(F^A_{QL} F^B_{QL})(F^C_{QL} F^D_{QL}),~~ {1 \over \Lambda_{PS}^2}\epsilon_{ABCD}(F^A_{u}F^B_{d})(F^C_{u}F^D_{d}) + \ldots,
    \label{eq:dim6}
\end{equation}
which set a lower bound on the cutoff scale $\Lambda_{PS} \gtrsim 10^{15} \text{ GeV}$. 

We use the same cutoff $\Lambda_{PS}$ for all the non-renormalizable operators in the Pati-Salam effective field theory.

\subsection{High-Scale Pati-Salam Breaking}
 \label{subsec:PSHigh}
When the Pati-Salam gauge group is broken at a high scale the type-I seesaw mechanism is an attractive way to generate large right-handed neutrino masses that then give rise to small left-handed Majorana neutrino masses. This is implemented by adding to the minimal particle content a field $\Delta$ in the $(\overline{10},1,-1)_{PS}$ representation. The interaction term $Y_{\nu}F_u^AF_u^B \Delta_{AB}$ generates the Majorana right-handed neutrino mass matrix $M_{\nu_R}$ when $\Delta$ gets a vev in its $\Delta_{44}$ component. The vev $\langle \Delta_{AB} \rangle = \delta_{A4}\delta_{B4}v_\Delta / \sqrt{2}$ is also responsible for breaking the Pati-Salam gauge group down to the SM. 

The heaviest active (left-handed) neutrino mass is $ \sim m_t^2 / m_{\nu_R}$, where we assumed $M^{{\rm Dirac}}_\nu \sim M_u$ (see eq.~\eqref{eq:masses}) and $m_{\nu_R}$ is the magnitude of the typical element in $M_{\nu_R}$.\footnote{We assume that all the elements in the right-handed neutrino masses are similar.} Observations of neutrino oscillations provide a lower bound on the heaviest active neutrino, which we take to be $\sqrt{\Delta m_\text{atm}^2} = 0.05$ eV~\cite{Esteban:2018azc}, while an upper bound arises from the cosmological limit on the sum of neutrino masses, $\sum_\nu m_\nu < 0.12$ eV~\cite{Aghanim:2018eyx}.\footnote{This is a 95 \% C.L. combined bound for Planck including TT, TE, EE, lowE, lensing, and baryonic acoustic oscillations, and it depends on the other parameters in the fit. For example, larger values of $H_0$ will give tighter constraints on this sum.}  These upper and lower bounds are not far apart and imply  that $m_{\nu_R} \sim 10^{14}$ GeV. To avoid strong coupling of the right-handed neutrinos to $\Delta$ this implies that $v_\Delta \gtrsim 10^{13}$ GeV. See ref.~\cite{Pati:2017ysg} for a related discussion of neutrino masses.

In sec.~\ref{sec:LQ} it was noted that in the context of the SM baryon number violating processes for $\Phi_{3}$ and $\Phi_{4}$ occur at dimension five. We now turn to studying the analogous fact when the SM is embedded in Pati-Salam. The leading contribution to a diquark coupling for the scalar leptoquarks now occurs at dimension six through terms involving $\Delta$. For example,
\begin{equation}
    \frac{1}{\Lambda_{PS}^2} (F_u^A F_d^B) (\Phi_{15}^\dagger)^C_E \, \, H^\dagger (\Delta^\dagger)^{ED}  \epsilon_{ABCD}  \ \ \stackrel{\langle \Delta \rangle}{\rightarrow} \ \ \frac{v_\Delta}{\sqrt{2}\Lambda_{PS}^2}(u_R^\alpha d_R^\beta)(\Phi_3^\dagger)^\gamma H^\dagger \epsilon_{\alpha \beta \gamma}.
    \label{eq:dim6Delta}
\end{equation}
We note that the same diquark coupling for $\Phi_3$ can be generated by either Higgs doublet ($H$ or $H_2$). The other case arises by replacing the $H$ boson field by the $\Phi_{15}$ field in the dimension-six Pati-Salam operator above as the following example shows, 
\begin{equation}
    \frac{1}{\Lambda_{PS}^2} (F_u^E  F_d^B) (\Phi_{15}^\dagger)_E^A  (\Phi_{15}^\dagger)^C_F (\Delta^\dagger)^{FD}  \epsilon_{ABCD}  \ \ \stackrel{\langle \Delta \rangle}{\rightarrow} \ \ \frac{v_\Delta}{\sqrt{2}\Lambda_{PS}^2}(u_R^\alpha d_R^\beta)(\Phi_3^\dagger)^\gamma H_2^\dagger \epsilon_{\alpha \beta \gamma}.
    \label{eq:dim6DeltaPhi}
\end{equation}
Either eq.~\eqref{eq:dim6Delta} or eq.~\eqref{eq:dim6DeltaPhi} lead to $\Delta B = - \Delta L = -1$ interactions through the following effective Hamiltonian at a scale much below $M_{\Phi_3}$,
\begin{equation}
    {\cal H}_\text{eff} = G_6 (\nu_{\mu L}^\dagger s_R^\alpha) (d_R^\beta u_R^\gamma) \epsilon_{\alpha \beta \gamma} + \text{h.c.}~,
    \label{eq:HeffKplus}
\end{equation}
where we have explicitly specified the flavor of quarks and leptons that will lead to proton decay via the dominant mode $p \to K^+ \nu_\mu$. From eq.~\eqref{eq:yukawaPhi3and4} and eq.~\eqref{eq:dim6Delta} it follows, after integrating out the $\Phi_3$, that the dimensionful Fermi-like parameter $G_6$ is given by
\begin{equation}
\label{c6}
    G_6 = \frac{v_\Delta v (Y_4^{22})^*}{ 2 \Lambda_{PS}^2 M_{\Phi_3}^2} \sim \frac{\sqrt{3} v_\Delta m_s(M_Z)}{2 \Lambda_{PS}^2M_{\Phi_3}^2},
\end{equation}
where in the right-hand side of the above equation the texture in eq.~\eqref{eq:texture} has been applied. We note that the fermion masses entering in the texture are defined at the quark-lepton unification scale. However there are other renormalizable effects that we have not included and so, for simplicity, we evaluate the fermion masses at the $Z$ boson mass. In this example we use $m_s(M_Z) = 55 \text{ MeV}$ (see for example ref.~\cite{Xing:2007fb}).

The decay width of proton through the mode $p \to K^+  \nu$ is given by
\begin{equation}
    \Gamma_{p \to K^+  \nu} = \frac{m_p}{32\pi}\left(1-\frac{m_{K^+}^2}{m_p^2}\right)^2 |G_6|^2 \beta_{K^+}^2 ,
    \label{eq:ptoKplusbarnu}
\end{equation}
where $\beta_{K^+} =  0.139 \text{ GeV}^2$~\cite{Aoki:2017puj} quantifies the hadronic matrix element $\langle K^+ | (du)_R s_R |p\rangle = \beta_{K^+} P_R \, u_p$. Here $u_p$ is the spinor of the proton. The experimental bound on the lifetime for the proton decay mode $\tau(p \to K^+ \nu) < 5.9 \times 10^{33} \text{ years}$~\cite{Abe:2014mwa} from the Super-Kamiokande collaboration imposes a lower bound of
\begin{equation}
    G_6 <1.9 \times 10^{-31}~ {\rm GeV}^{-2}.
\end{equation}

This implies through eq.~\eqref{c6} that the vev of $\Delta$ satisfies that
\begin{equation}
    v_\Delta \lesssim 1 \times 10^{15} \text{ GeV} \left(\frac{M_{\Phi_3}}{2\text{ TeV}}\right)^2 \left(\frac{\Lambda_{PS}}{M_{PL}}\right)^2.
    \label{eq:vDelta}
\end{equation}
Even for $\Lambda_{PS} = M_{PL}$ the above bound is barely large enough to accommodate the right-handed neutrino masses without strong coupling. Avoiding strong coupling for the right-handed neutrino interactions with $\Delta$ together within the above bound on $v_{\Delta}$ gives, 
\begin{equation}
    \Lambda_{PS} \gtrsim 10^{18} \text{ GeV} \left(\frac{2 \text{ TeV}}{M_{\Phi_3}}\right),
    \label{eq:LambdaPSPlanck}
\end{equation}
which is only one order of magnitude away from the Planck scale. 

The right-hand sides of eqs.~\eqref{eq:vDelta} and~\eqref{eq:LambdaPSPlanck} are strongly impacted by the smallness of the Yukawa coupling $Y_4^{22} \sim 3\times 10^{-4}$. This factor arises because with the minimal model we are considering this coupling is related to the strange quark mass.

The bounds in eqs.~\eqref{eq:vDelta} and~\eqref{eq:LambdaPSPlanck} will be improved by future data from the Hyper-Kamiokande~\cite{Abe:2018uyc} and DUNE~\cite{Acciarri:2015uup} collaborations.

With the minimal particle content the renormalizable Pati-Salam couplings have a ${\rm U}(1)$ fermion number symmetry. This is broken down to a discrete $Z_4$ subgroup by the non-renormalizable four-fermion operators in eq.~\eqref{eq:dim6}. This means that quantum corrections cannot generate from these dimension-six operators the dimension-five baryon number violating couplings of the leptoquark scalars mentioned in section~\ref{sec:LQ} since they only respect a $Z_2$ subgroup of this $Z_4$. However this is no longer true when one adds fields to the minimal Pati-Salam particle content that generate neutrino masses.

\begin{figure}[t]
\begin{equation*}
\begin{gathered}
\begin{tikzpicture}[line width=1.5 pt,node distance=1 cm and 1 cm]
\coordinate[label=left:$F_u \supset u_R$](p1);
\coordinate[below right =of p1](v1);
\coordinate[above right = 1.42 cm of v1](mN);
\coordinate[below right = 1.42 cm of v1](H);
\coordinate[below  = 0.85 cm of H](Hdown);
\coordinate[below  = 0.45 cm of H](Hdown2);
\coordinate[above  = 1.15 cm of v1,label=above:$\cup$](nuR1);
\coordinate[above  = 1.5cm of v1,label=above:$F_u$](nuR2);
\coordinate[left = 0.65 cm of mN,label=left:$\nu_R$](nuRleft);
\coordinate[right = 0.65 cm of mN,label=right:$\nu_R \subset F_u$](nuRleft);
\coordinate[left = 0.65 cm of H,label=left:$d_R$](dRleft);
\coordinate[left = 0.65 cm of Hdown,label=left:$F_d \,$](dRleft);
\coordinate[left = 0.75 cm of Hdown2,label=left:$\cap$](dRleft);
\coordinate[right = 0.65 cm of H,label=right:$Q_L \subset F_{QL}$](Qright);
\coordinate[above = 0.7 cm of mN,label=above:$\langle \Delta \rangle $](mNE);
\coordinate[below left = of v1,label=left:$F_d \supset d_R$](p2);
\coordinate[right = 2 cm of v1](aux);
\coordinate[right = 1.75 cm of v1](aux1);
\coordinate[right = 1 cm of aux,label=right:$\Phi_3 \subset \Phi_{15}$](p3E);
\coordinate[below = 0.7 cm of H,label=below:$H$](p4E);
\draw (mN) node[cross=5pt,rotate=0,black]{};
\draw[fermion] (p1)--(v1);
\draw[fermion] (p2)--(v1);
\draw[scalarnoarrow] (mN)--(mNE);
\draw[scalar] (aux)--(p3E);
\draw[scalar] (H)--(p4E);
\draw[fermion] (p2)--(v1);
\draw[fill=black] (aux) circle (0.075cm);
\draw[fill=black] (H) circle (0.075cm);
\semiloop[fermionD]{v1}{aux}{0}
\semiloop[fermionbar]{v1}{aux}{0}
\semiloopI[fermionD]{v1}{aux}{0}
\semiloopI[fermionbar2]{v1}{aux}{0}
\draw[fill=gray] (v1) circle (0.15cm);
\end{tikzpicture}
\end{gathered}
\quad \quad  
\begin{gathered}
\begin{tikzpicture}[line width=1.5 pt,node distance=1 cm and 1 cm]
\coordinate[label=left:$F_u \supset u_R$](p1);
\coordinate[below right =of p1](v1);
\coordinate[above right = 1.42 cm of v1](mN);
\coordinate[below right = 1.42 cm of v1](H);
\coordinate[below  = 0.85 cm of H](Hdown);
\coordinate[below  = 0.45 cm of H](Hdown2);
\coordinate[above  = 1.15 cm of v1,label=above:$\cup$](nuR1);
\coordinate[above  = 1.5cm of v1,label=above:$F_u$](nuR2);
\coordinate[left = 0.65 cm of mN,label=left:$\nu_R$](nuRleft);
\coordinate[right = 0.65 cm of mN,label=right:$\nu_R \subset F_u$](nuRleft);
\coordinate[left = 0.65 cm of H,label=left:$d_R$](dRleft);
\coordinate[left = 0.65 cm of Hdown,label=left:$F_d \,$](dRleft);
\coordinate[left = 0.75 cm of Hdown2,label=left:$\cap$](dRleft);
\coordinate[right = 0.65 cm of H,label=right:$L_L \subset F_{QL}$](Qright);
\coordinate[above = 0.7 cm of mN,label=above:$\langle \Delta \rangle $](mNE);
\coordinate[below left = of v1,label=left:$F_d \supset d_R$](p2);
\coordinate[right = 2 cm of v1](aux);
\coordinate[right = 1.75 cm of v1](aux1);
\coordinate[right = 1 cm of aux,label=right:$H$](p3E);
\coordinate[below = 0.7 cm of H,label=below:$\qquad \ \quad \Phi_3\subset \Phi_{15}$](p4E);
\draw (mN) node[cross=5pt,rotate=0,black]{};
\draw[fermion] (p1)--(v1);
\draw[fermion] (p2)--(v1);
\draw[scalarnoarrow] (mN)--(mNE);
\draw[scalar] (aux)--(p3E);
\draw[scalar] (H)--(p4E);
\draw[fermion] (p2)--(v1);
\draw[fill=black] (aux) circle (0.075cm);
\draw[fill=black] (H) circle (0.075cm);
\semiloop[fermionD]{v1}{aux}{0}
\semiloop[fermionbar]{v1}{aux}{0}
\semiloopI[fermionD]{v1}{aux}{0}
\semiloopI[fermionbar2]{v1}{aux}{0}
\draw[fill=gray] (v1) circle (0.15cm);
\end{tikzpicture}
\end{gathered}
\end{equation*}
\caption{Examples of one-loop contributions to the diquark coupling of $\Phi_3$ in eq.~\eqref{eq:loop}.}
\label{fig:loop}
\end{figure}
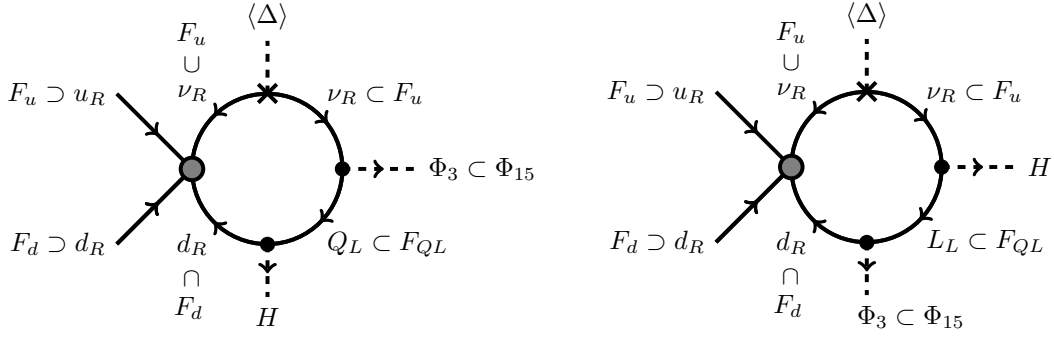

As fig.~\ref{fig:loop} shows, in this case the diquark coupling for the leptoquark $\Phi_3$ can also be generated at the quantum level through, for example, the one-loop diagrams displayed in this figure.
The contribution of these loop diagrams to the diquark coupling of $\Phi_3$ is
\begin{equation}
  \sim \frac{1}{16 \pi^2} {1  \over \Lambda_{PS}^2}\left[ (Y_2^{\nu 3})^*Y_3^{33} + Y_4^{33}(Y_1^{3\nu})^*\right]{\rm log}\left({\Lambda_{PS}^2 \over M_{\Phi_3}^2} \right) \epsilon_{\alpha \beta \gamma} m_{\nu_R}  (d_R^{\alpha} u_R^{\beta}) (\Phi_3^\dagger)^\gamma   H^\dagger,
 \label{eq:loop} 
\end{equation} 
where we have assumed the heaviest quarks are in the loop.
The term on the left of the expression between squared brackets in eq.~\eqref{eq:loop} refers to the contribution from the loop process in the left panel of fig.~\ref{fig:loop}, while the term on the right corresponds to the contribution from the loop process displayed in the right panel of the same figure. 
The logarithm indicates that this should be thought of as an operator mixing effect. From the scale $\Lambda_{PS}$ to the scale of Pati-Salam symmetry breaking it is mixing between Pati-Salam invariant dimension-six operators and below the scale of Pati-Salam breaking it is mixing between the dimension-six baryon number violating four-fermion operators in the SM and the effectively dimension-six operator $\epsilon_{\alpha \beta \gamma} m_{\nu_R}  (d_R^{\alpha} u_R^{\beta}) (\Phi_3^\dagger)^\gamma  H^\dagger$. This mixing effect does not give a stronger lower bound on $\Lambda_{PS}$ than eq.~\eqref{eq:LambdaPSPlanck}.

\subsection{Low-Scale Pati-Salam Breaking}
\label{subsec:PSLow}
For Pati-Salam breaking at the low scale, we use the inverse seesaw mechanism to generate neutrino masses~\cite{Perez:2013osa}. The Pati-Salam gauge group is broken through the vev of the field $\chi \sim (4,1,1/2)_{PS}$, explicitly $\langle \chi^A \rangle = \delta^{A4} v_\chi /\sqrt{2}$. The field $\Delta$ introduced earlier does not occur in the low-scale breaking scenario.

In the Pati-Salam theory there are no dimension-five or dimension-six operators whose tree-level matrix elements give rise to the dimension-five diquark couplings of $\Phi_{3}$ and $\Phi_{4}$ discussed in Section~\ref{sec:LQ}. Baryon number violation involving these leptoquarks first occurs at dimension seven through the following interactions,\footnote{Note that analogous operators exist with the $H$ replaced by $H_2$ in $\Phi_{15}$. For example, in eq.~\eqref{eq:Phi3dim7}, $(F_u^E  F_d^B)(\Phi_{15}^\dagger)_E^A(\Phi_{15}^\dagger \chi)^C \chi^D \epsilon_{ABCD} / \Lambda_{PS}^3$.}
\begin{eqnarray}
  &\frac{1}{\Lambda_{PS}^3} (F_{u}^{A} F_{d}^{B}) (\Phi_{15}^\dagger \chi)^C \chi^D H^\dagger  \epsilon_{A B C D}  \quad \quad &\stackrel{\langle \chi \rangle}{\rightarrow}  \quad \frac{v_\chi^2}{\Lambda_{PS}^3}(d_R^\alpha u_R^\beta) (\Phi_3^\dagger)^\gamma  H^\dagger \epsilon_{\alpha \beta \gamma},  \label{eq:Phi3dim7}\\
 & \frac{1}{\Lambda_{PS}^3} (F_{QL}^A F_{QL}^B) (\Phi_{15}^\dagger \chi)^C \chi^D H^\dagger \epsilon_{ABCD} \quad &\stackrel{\langle \chi \rangle}{\rightarrow} \quad \frac{v_\chi^2}{\Lambda_{PS}^3}(Q_L^\alpha Q_L^\beta) (\Phi_3^\dagger)^\gamma H^\dagger \epsilon_{\alpha \beta \gamma}, \label{eq:dim7Low1}\\
 &\frac{1}{\Lambda_{PS}^3} (F_{d}^{A} F_{d}^{B}) (\Phi_{15}^\dagger \chi)^C  \chi^D H \epsilon_{A B C D} \quad \quad \ \, &\stackrel{\langle \chi \rangle}{\rightarrow} \quad \frac{v_\chi^2}{\Lambda_{PS}^3} (d_R^\alpha d_R^\beta) (\Phi_3^\dagger)^\gamma H \epsilon_{\alpha \beta \gamma}, \label{eq:dim7Low2}\\
  &\frac{1}{\Lambda_{PS}^3} (F_{d}^{A} F_{d}^B) (\Phi_{15} \chi)^C \chi^D H^\dagger \epsilon_{ABCD} \quad \quad \ \, &\stackrel{\langle \chi \rangle}{\rightarrow} \quad \frac{v_\chi^2}{\Lambda_{PS}^3} (d_R^\alpha d_R^\beta) \Phi_4^\gamma H^\dagger \epsilon_{\alpha \beta \gamma}.
  \label{eq:dim7Low3}
 \end{eqnarray}
 and 
 \begin{eqnarray}
 &\frac{1}{\Lambda_{PS}^3} (F_u^A F_d^B) (\Phi_{15}^\dagger \chi)^C (\Phi_{15}^\dagger \chi)^D \epsilon_{ABCD} \quad  & \stackrel{\langle \chi \rangle}{\rightarrow} \quad \frac{v_\chi^2}{\Lambda_{PS}^3} (u_R^\alpha e_R) (\Phi_3^{\dagger})^\beta (\Phi_3^\dagger)^\gamma \epsilon_{\alpha \beta \gamma}, \label{eq:dim7Low4}\\
 &\frac{1}{\Lambda_{PS}^3} (F_{QL}^A F_{QL}^B) (\Phi_{15}^\dagger \chi)^C (\Phi_{15}^\dagger \chi)^D \epsilon_{ABCD} \quad  & \stackrel{\langle \chi \rangle}{\rightarrow} \quad \frac{v_\chi^2}{\Lambda_{PS}^3} (Q_L^\alpha L_L) (\Phi_3^{\dagger})^\beta (\Phi_3^\dagger)^\gamma \epsilon_{\alpha \beta \gamma},\label{eq:dim7Low5} \\
 &\frac{1}{\Lambda_{PS}^3}(F_d^A F_d^B) (\Phi_{15}^\dagger \chi)^C (\Phi_{15} \chi)^D \epsilon_{ABCD} \quad & \stackrel{\langle \chi \rangle}{\rightarrow} \quad  \frac{v_\chi^2}{\Lambda_{PS}^3}(d_R^\alpha e_R) (\Phi_3^\dagger)^\beta \Phi_4^\gamma \epsilon_{\alpha \beta \gamma}. \label{eq:dim7Low6}
 \end{eqnarray}
 Here for example $(\Phi_{15}^\dagger \chi)^C = (\Phi_{15}^{\dagger})^C_F \chi^F$.
 Eqs.~\eqref{eq:Phi3dim7},~\eqref{eq:dim7Low1},~\eqref{eq:dim7Low2}, and \eqref{eq:dim7Low3} are analogous to the non-renormalizable operators in eq.~\eqref{eq:dim5SLQ}, while eqs.~\eqref{eq:dim7Low4},~\eqref{eq:dim7Low5}, and~\eqref{eq:dim7Low6} are analogous to those in eq.~\eqref{eq:dim5SLQextra}.
 Note that the leptoquark couplings above are proportional to $v_\chi^2 / \Lambda_{PS}^3$. Recall that an acceptable nucleon decay rate requires $\Lambda_{PS} > 10^{15}$ GeV. Therefore, since we are considering low-scale breaking $v_\chi \sim 100-1000$ TeV, the dimension-seven operators above will not give rise to observable baryon number violation. 
 
 Vector leptoquarks with the same SM quantum numbers $(3,1,2/3)_{SM}$ as the broken generators of Pati-Salam give rise to baryon number violation at dimension five in the SM effective field theory~\cite{Assad:2017iib}. However, in Pati-Salam models
 their effects are much more suppressed. Baryon number violation does not occur until dimension seven in the Pati-Salam effective theory. For example, through the operator
 \begin{equation}
 \frac{1}{\Lambda_{PS}^3}  (F_{QL}^A  \sigma_\mu F_d^B) \chi^C H^\dagger (D^\mu \chi)^D  \epsilon_{ABCD}.
 \end{equation}
Similarly to the case of the leptoquarks $\Phi_3$ and $\Phi_4$, the gauge bosons associated with the broken generators of ${\rm SU}(4)$ will not give rise to unacceptably large baryon number violation.

\subsection{Additional Particle Content}
In principle the theory could have a larger scalar content than the minimal scenarios presented so far. Here we aim to illustrate with a simple example how the symmetry of the Pati-Salam theory can protect mixed scalars (with both leptoquark and diquark couplings at the renormalizable level) with masses at the TeV scale that would be naively ruled out from the low-energy perspective. For example, add to the minimal content a single scalar field with the following quantum numbers
\begin{equation}
\Phi_{10} = \begin{pmatrix} S_\text{DQ} & S_\text{LQ} \\ S_\text{LQ} & \delta^{++} \end{pmatrix} \sim (\overline{10},1,1)_{PS}.
\end{equation}
This symmetric ${\rm SU}(4)$ representation contains a scalar leptoquark $S_\text{LQ} \sim (\bar 3,1, 4/3)_{SM}$, a scalar diquark $S_\text{DQ} \sim (\bar 6, 1, 2/3)_{SM}$, and a doubly charged scalar $\delta^{++} \sim (1,1,2)_{SM}$. The new Yukawa interactions are given by
\begin{equation}
\begin{split}
   - {\cal L} &= Y_{10} (F_d^A F_d^B) (\Phi_{10})_{AB}  + \text{h.c.}\\
    &=Y_{10} (d_R^\alpha d_R^\beta) (S_{\text{DQ}})_{\alpha \beta} + 2Y_{10} (d_R^\alpha e_R) (S_{\text{LQ}})_\alpha + Y_{10} (e_R e_R) \delta^{++} + \text{h.c.}~,
    \end{split}
    \label{eq:renormalizable10}
\end{equation}
where $Y_{10}$ is symmetric in the flavor space.  Already at the renormalizable level we can see that, contrary to what we expected from section~\ref{sec:LQ}, the scalar $S_\text{LQ}$ is a pure leptoquark in this case. The ${\rm SU}(4)$ symmetry forbids the diquark coupling of $S_\text{LQ}$ with matter, that would be otherwise allowed by the SM gauge symmetry and would rule out a $S_\text{LQ}$ lighter than $10^{12}$ GeV (for order one Yukawa couplings). 

To proceed further we first consider the case of high-scale breaking. We imagine that only the $S_\text{LQ}$ scalar is allowed to be at the TeV scale.
A diquark coupling for $S_\text{LQ}$ can be generated at dimension six through the following non-renormalizable interaction:
\begin{equation}
\frac{1}{\Lambda_{PS}^2}(F_u^A F_u^B) (\Phi_{10}^\dagger)^{CF}\Delta_{FG} (\Delta^\dagger)^{GD} \epsilon_{ABCD} + \text{h.c.}  \ \ \stackrel{\langle \Delta \rangle}{\rightarrow}  \ \ \frac{v_\Delta^2}{\Lambda_{PS}^2} (u_R^\alpha  u_R^\beta)  (S_\text{LQ}^\dagger)^\gamma \epsilon_{\alpha \beta \gamma} + \text{h.c.}~.
\label{eq:Delta10dim6}
\end{equation}
Note that the above operator is antisymmetric in the flavor indices and, as discussed in section~\ref{sec:LQ}, it can only mediate proton decay by the exchange of a $W$ gauge boson as the top-left panel of fig.~\ref{fig:FG} illustrates. Identifying the diquark coupling in eq.~\eqref{eq:Wexchange} as $Y_\text{DQ} = v_\Delta^2 / \Lambda_{PS}^2$ and the leptoquark coupling as $Y_\text{LQ} = 2Y_{10}^{11}$, the bound from eq.~\eqref{eq:Wexchange} can be translated to the operator in the right-hand side of eq.~\eqref{eq:Delta10dim6}, giving
\begin{equation}
v_\Delta \lesssim 10^{10} \text{ GeV} \left(\frac{M_{S_\text{LQ}}}{2\text{ TeV}}\right)\left(\frac{\Lambda_{PS}}{M_{PL}}\right),
\label{eq:SLQres}
\end{equation}
where $Y_{10}^{11} \simeq 1$ has been assumed. This rules out the high-scale breaking scenario under the assumption of an order one Yukawa coupling for a TeV scale leptoquark $S_\text{LQ}$ since $v_\Delta \gtrsim 10^{13}$ GeV is required to accommodate the active neutrino masses.

Next we consider low-scale breaking. In this case the diquark coupling of $S_\text{LQ}$ arises from the following dimension-six Pati-Salam invariant operator
\begin{equation}
\frac{1}{\Lambda_{PS}^2} (F_u^A F_u^B) (\Phi_{10}^\dagger \chi^\dagger)^C \chi^D \epsilon_{ABCD} + \text{h.c.} \ \ \stackrel{\langle \chi \rangle}{\rightarrow}  \ \ \frac{v_\chi^2}{\Lambda_{PS}^2} (u_R^\alpha  u_R^\beta)  (S_\text{LQ}^\dagger)^\gamma \epsilon_{\alpha \beta \gamma} + \text{h.c.}~.
\label{eq:diquark10}
\end{equation}
Now eq.~\eqref{eq:SLQres} holds when one replaces the vev of $\Delta$ by the vev of $\chi$. 
Note that if $\Lambda_{PS} \simeq 10^{15}$ GeV and $M_{S_\text{LQ}} \simeq 2 $ TeV, the bound becomes $v_\chi \lesssim 1000 \text{ TeV}$ which is near the limit imposed by the decay $K_L \to \mu^\pm e^\mp$.

With low scale breaking the scalars $\Phi_3$, $\Phi_4$ and $S_\text{LQ}$ are all light. Baryon number violation occurs at dimension-five  through the operator\footnote{Note that the left-hand side of eq.~\eqref{eq:alllight} can also lead to other $|\Delta B| = 1$ processes by contracting the ${\rm SU}(4)$ indices differently. The bounds derived in those cases are very similar to the example in the right-hand side of eq.~\eqref{eq:alllight}.}
\begin{equation}
 \frac{1}{\Lambda_{PS}} (\Phi_{10}^\dagger)^{A F} (\Phi_{15})_F^B(\Phi_{15}^\dagger)^C_E \chi^E \chi^D \epsilon_{ABCD} \  \stackrel{\langle \chi \rangle }{\to}  \ \frac{v_\chi^2}{\Lambda_{PS}} (S_\text{LQ}^\dagger)^\alpha \Phi_4^\beta (\Phi_3^\dagger)^\gamma  \epsilon_{\alpha \beta \gamma} .
 \label{eq:alllight}
\end{equation}
Such operator, together with the renormalizable interactions of the scalar leptoquarks $\Phi_3$ and $\Phi_4$ given in eq.~\eqref{eq:yukawaPhi3and4}, and the renormalizable interactions of the scalar leptoquark $S_{LQ}$ given in eq.~\eqref{eq:renormalizable10}
give rise to $\Delta B = -\Delta L = -1$ processes such as $n \to \nu e^+ e^-$ through an effective Hamiltonian
\begin{equation}
 {\cal H}_\text{eff} =  \frac{v_\chi^2}{\Lambda_{PS}} \frac{Y_{10}^{11}|Y_{4}^{11}|^2}{M_{S_{LQ}}^2 M_{\Phi_3}^2 M_{\Phi_4}^2} (u_L^\alpha e_R^\dagger) (\nu_L^\dagger d_R^\beta)(d_R^\gamma e_R) \epsilon_{\alpha \beta \gamma} + \cdots~ + \text{h.c.}~.
  \label{eq:Hamiltonian9}
\end{equation}
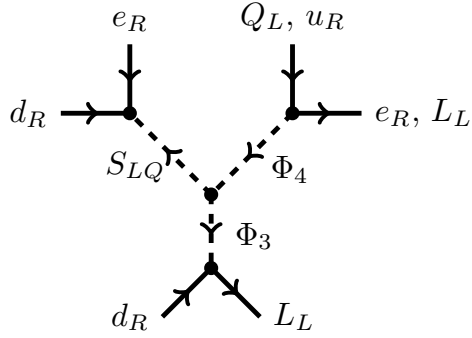
\begin{figure}
    \centering
\begin{equation*}
\begin{gathered}
\resizebox{.4\textwidth}{!}{
\begin{tikzpicture}[line width=1.5 pt,node distance=1 cm and 1.5 cm]
\coordinate(vcentral);\coordinate[above left =  1.25cm of vcentral](v1);
\coordinate[above right =  1.25cm of vcentral](v2);\coordinate[below  = 0.8cm of vcentral](v3);
\coordinate[below =  0.9 cm of v1,label=$\ S_{LQ}$](v1aux);\coordinate[below =  0.9 cm of v2,label=$\! \Phi_4$](v2aux);\coordinate[below =  0.45 cm of vcentral,label=right:$\ \Phi_3$](v3aux);
\coordinate[left= 0.75 cm of v1,label=left:$d_R$](f1a);
\coordinate[above = 0.75cm of v1,label=above:$e_R$](f1b);
\coordinate[above = 0.75cm of v2,label=above:$Q_L\text{, }u_R$](f2a);
\coordinate[right = 0.75 cm of v2,label=right:$e_R\text{, }L_L$](f2b);
\coordinate[below right = 0.75 cm of v3,label=right:$L_L$](f3a);\coordinate[below left = 0.75 cm of v3,label=left:$d_R$](f3b);
\draw[fermion](f1a)--(v1);
\draw[fermion](f1b)--(v1);
\draw[fermion](f2a)--(v2);
\draw[fermion](v2)--(f2b);
\draw[scalar] (vcentral) -- (v1);
\draw[scalar] (v2) -- (vcentral);
\draw[scalar] (vcentral) -- (v3);
\draw[fermion] (v3)--(f3a);
\draw[fermion] (f3b) -- (v3);
\filldraw[fill=black] (vcentral) circle (.05cm);\draw[fill=black] (v1) circle (.05cm);
\draw[fill=black] (v2) circle (.05cm);
\draw[fill=black] (v3) circle (.05cm);
\end{tikzpicture}}
\end{gathered} 
\end{equation*}
    \caption{Diagram for $\Delta B = -\Delta L =-1$ processes.}
    \label{fig:MixedLight}
\end{figure}
Fig.~\ref{fig:MixedLight} shows the diagram for this process. Following ref.~\cite{Hambye:2017qix}, the bound on the vev of $\chi$ is given by
\begin{equation}
v_\chi \lesssim 3 \times 10^{10} \text{ GeV}\left( \frac{M_{S_\text{LQ}}}{2\text{ TeV}} \right)\left(\frac{M_{\Phi_3}}{2\text{ TeV}} \right) \left(\frac{M_{\Phi_4}}{2\text{ TeV}} \right) \left(\frac{\Lambda_{PS}}{M_{PL}}\right)^{1/2},
\label{eq:vChimu}
\end{equation}
where we used $Y_4^{11} \simeq m_d(M_Z)/v \simeq 1.7 \times 10^{-5}$, and assumed that $Y_{10}^{11} \simeq 1$. Here the bound on the lifetime $\tau(n \to  \nu e^+ e^-)$~\cite{McGrew:1999nd} from the IMB-3 experiment has been used. The bound in eq.~\eqref{eq:vChimu} is influenced by the smalness of the Yukawa couplings for the leptoquarks $\Phi_3$ and $\Phi_4$. If they were taken to be order one, the above bound would be five orders of magnitude stronger. 
 
 Our results are strongly dependent on the field content we have assumed. It is well-known that renormalizable baryon number violation is possible in Pati-Salam models~\cite{Pati:1983zp}. Consider for example adding to the minimal theory based on Pati-Salam symmetry a scalar field 
  \begin{equation}
    \Phi_6 = \begin{pmatrix} \epsilon^{\alpha \beta \gamma} (\phi_\text{DQ})_\gamma && \phi_\text{LQ}^\alpha \\ - \phi_\text{LQ}^\alpha && 0 \end{pmatrix}\sim (6,1,0)_{PS},
 \end{equation}
 where $\phi_\text{DQ} \sim (\bar 3,1,1/3)_{SM}$ and $\phi_\text{LQ} \sim (3,1,-1/3)_{SM}$. According to table~\ref{tab:SLQ}, $\phi_\text{LQ}$ and $\phi_\text{DQ}$ are mixed scalars. 
 This is also true in Pati-Salam models as can be seen explicitly in their Yukawa couplings,
 \begin{eqnarray}
 -{\cal L} &=&   Y_6 \, F_u^AF_d^B (\Phi_6^\dagger)_{AB} + Y_6' \, \epsilon_{ABCD}F_u^A F_d^B \Phi_6^{CD} \nonumber\\
 &&+ Y_6'' F_{QL}^AF_{QL}^B (\Phi_6^\dagger)_{AB} 
 + Y_6''' \epsilon_{ABCD} F_{QL}^A F_{QL}^B \Phi^{CD}_6  +  \text{h.c.}  \\
  &=&  Y_6 \epsilon_{\alpha \beta \gamma} u_R^\alpha d_R^\beta (\phi_\text{DQ}^\dagger)^\gamma +2 Y_6' u_R^\alpha e_R  \, \phi_{DQ\alpha} + \cdots + \text{h.c.}~.
 \end{eqnarray}
This example shows that even after the embedding of $\phi_\text{DQ}$ and $\phi_\text{LQ}$ in a Pati-Salam model through $\Phi_6$, they give rise to renormalizable baryon number violation.

\section{Concluding Remarks}
Treating the SM as an effective theory it contains higher dimension operators that give rise to baryon number violation at dimension six. There are a number of flavor anomalies that can be addressed by adding to the SM scalar leptoquarks with masses around a few TeV. However, adding such leptoquarks to the SM effective theory one finds that baryon number violation can now occur at the renormalizable and/or dimension-five levels. Experimental limits on baryon number violating processes force the  leptoquarks mass to be too large to  explain the flavor anomalies even when the cutoff of the SM effective theory is the Planck mass.

To address this issue, we have studied embedding the SM in the minimal Pati-Salam model with one or more leptoquark scalars with masses at the TeV scale. The Pati-Salam model was treated as an effective theory with a cutoff scale $\Lambda_{PS} \le M_{PL}$ that suppresses higher dimension operators. 

When the Pati-Salam model is broken at a high scale and the neutrino masses arise from the type-I seesaw mechanism we found by using experimental constraints on baryon number violation that $\Lambda_{PS} \gtrsim 10^{18} \text{ GeV}$ and the vev defining the Pati-Salam breaking is constrained  to be in the window $10^{13} \lesssim v_\Delta \lesssim 10^{15}$ GeV. 

When the Pati-Salam model (with minimal particle content) is broken at the low scale, neutrinos get mass through the inverse seesaw mechanism. In this case in the Pati-Salam effective field theory baryon number violating operators involving the leptoquarks first occur at dimension seven. Baryon number violation arising from dimension-six operators involving four fermions will dominate in this case and provide a bound $\Lambda_{PS} \gtrsim 10^{15}$ GeV.

We also studied a case where the Pati-Salam model had non-minimal particle content. Adding a scalar to the SM that has both leptoquark and diquark couplings gives rise to renormalizable baryon number violation through the tree-level exchange of that scalar. We showed in an example that embedding the SM and such a scalar in the Pati-Salam effective field theory can protect the theory for having unacceptably large baryon number violation. 

It is worth emphasizing that our conclusions depend on the field content of the model. Adding more degrees of freedom, several of which are at the TeV scale, can lead to unacceptably large baryon number violation.

An interesting extension of this work would be to study how other embeddings of the SM with additional TeV scale leptoquarks influences baryon number violation. 

\begin{acknowledgments}
We thank Pavel Fileviez P\'erez for helpful discussions at the beginning of this project. We thank Bartosz Fornal and Alexis D. Plascencia for several helpful comments.
This material is based upon work supported by the U.S. Department of Energy, Office of Science, Office of High Energy Physics, under Award Number DE-SC0011632 and by the Walter Burke Institute for Theoretical Physics.
\end{acknowledgments}

\bibliography{leptoquarksPS}

\end{document}